\documentclass[12pt,twoside]{article}

\begin{document}

{\bf What is "system": the information-theoretic arguments}

\bigskip

\bigskip

\centerline{\it M. Dugi\' c$^{\ast}$, J. Jekni\' c-Dugi\'
c$^{\dag}$}

\medskip

$^{\ast}${Department of Physics, Faculty of Science, Kragujevac,
Serbia\\
 E-mail: dugic@kg.ac.yu}

$^{\dag}${Department of Physics, Faculty of Science, Ni\v s,
Serbia\\
 E-mail: jjeknic@ni.ac.yu}

\bigskip

{\bf Abstract:} The problem of "what is 'system'?" is in the very
foundations of modern quantum mechanics. Here, we point out the
interest in this topic in the information-theoretic context. E.g.,
we point out the possibility to manipulate a pair of mutually
non-interacting, non-entangled systems to employ entanglement of
the newly defined "(sub)systems" consisting the one and the same
composite system. Given the different divisions of a composite
system into "subsystems", the Hamiltonian of the system may
generate in general non-equivalent quantum computations.
Redefinition of "subsystems" of a composite system may be regarded
as a method for avoiding decoherence in the quantum hardware. In
principle, all the notions refer to a composite system as simple
as the hydrogen atom.

\bigskip

\bigskip

{\bf 1. Introduction}

\bigskip

\noindent A physical system is defined by its degrees of freedom
(and, usually, by the conjugate momenta) and by the system's
parameters (such as the mass, electric charge etc.). While this
reasoning is taken for granted in the classical physics domain, it
is not so in the context of quantum mechanics. E.g.--as Bohr
pointed it out--quantum mechanics does not in general allow the
{\it a priori} introduction (definition) of the system
observables, that elevates almost to a paradox for the complex
systems. Actually, quantum entanglement--typical for complex
systems--prevents us from determining the states of the entangled
subsystems. On the other side, however, without the possibility
to pose (in the classical-physics--like manner) a border line
between the (sub)systems, one can not even pose certain quantum
mechanical problems (such as e.g. the measurement problem) [1].
Therefore, there is a strong need for a method determining the
system's observables and the border line between the systems.

The problem of "what is 'system'?" stems from our classical
intuition that deals with {\it objects}, and it seems natural to
seek for an answer to the problem in the context of the
decoherence theory--which is widely believed to be the missing
link between the "quantum" and "classical" [1, 2,  3, 4].
Following partly the reasoning of Ref. [1], an operational method
for defining a subsystem is proposed in Ref. [5] that is based on
the foundations of the decoherence theory [6, 7]. Fortunately
enough, targeting equally the issue of the entanglement formation
[8], the method of Ref. [5] may be applied even for the isolated
systems. The main observation of Ref. [5] is the {\it relativity
of the concept of physical system}. And this is the starting point
of the present paper.

The system relativity assumes a redefinition of a complex system
in terms of certain new subsystems: the {\it canonical
transformations} of the degrees of freedom (and, in general, of
the conjugate momenta--cf. Appendix A) of the subsystems
consisting the composite system may introduce the new subsystems,
thus implementing the question "what is '(sub)system'?" [5]. To
this end, the re-arrangement (or regrouping) of the subsystems of
a composite system--typical for quantum measurement theory--as
well as introducing the new variables of the unique subsystem are
merely trivial (relative to the general (linear) canonical
transformations) and will not be considered in this paper. As
defined by the canonical transformations, the new "subsystems"
{\it need not bear any intuitive contents} as "physical
(sub)systems" (cf. Appendix A).

In this paper, we make slightly a turn in our perspective to the
problem. Actually, we give an information-theoretic re-formulation
of the problem at issue by starting from the following plausible
assumption: any reasonable definition of "system" should be based
on certain {\it information} (about "system"). An information
acquired by measurement distinguishes accessibility of the
measurement i.e. accessibility of certain system's observables.
And these are exactly those observables that might constitute a
definition of the system.

As an output of our considerations, we point out that the
relativity of the concept of "system" makes some of the
aforementioned, plausible notions also to be relative; i.e., one
should always {\it take care about the actual division} of a
composite system into subsystems. Depending on the answer to the
question of which division is in question, it is plausible to
expect that the answers would not mutually be equivalent. E.g.,
neither entanglement nor quantum-computation-routes need to be
equivalent for the different divisions of the composite system
into subsystems. Interestingly enough, the method for definimg
subsystems may be considered as a method for avoiding decoherence
in the quantum hardware. Finally, we briefly emphasize the
relevance of our general considerations for the {\it realistic}
physical models.

\bigskip

{\bf 2. Hydrogen atom as an isolated quantum system}

\bigskip

\noindent

\noindent Paradigmatic for our discussion is a system as simple as
the hydrogen atom. Actually, we are concerned with a bipartite
quantum system $\mathcal{C}$ consisting of the two subsystems,
$\mathcal{A}$ and $\mathcal{B}$ ($\mathcal{C} = \mathcal{A} +
\mathcal{B}$). Let us assume that the subsystems are not in any
mutual interaction. Then, it is generally assumed that such a
system can {\it not} be {\it directly} used for quantum
information processing as long as entanglement is required for
information processing.

However, as we show in the sequel, this is not necessarily the
case. Actually, a proper definition of the {\it new subsystems}
(cf. Ref. [5]), $\mathcal{E}$ and $\mathcal{F}$ of $C$
($\mathcal{C} = \mathcal{E} + \mathcal{F}$), may help in providing
entanglement without introducing any interaction between
$\mathcal{A}$ and $\mathcal{B}$ as well as without any further
operations exerted locally on $\mathcal{A}$ and/or $\mathcal{B}$.
This observation comes {\it directly} from the {\it relativity of}
the concept of physical {\it system} [5].

In order to make our discussion more intuitive, we shall mainly
refer in this section to the model of the hydrogen atom as an
isolated quantum system. Let us suppose that the system "electron
+ proton ($e + p$ )" is an isolated quantum system out of any
external field. The standard theory of the hydrogen atom
(neglecting spin) gives rise to a redefinition of the atom as it
is well known: "center-of-mass of the atom + the relative
particle ($CM + R$)"--cf. eqs. (4), (5) as the special case of
the general transformations as defined in Appendix A. While there
is the Coulomb interaction in the $e + p$ system, there is not any
interaction in the system $CM + R$. The later allows one exactly
to solve the problem of the internal energies of the atom by the
variables separation. Apart from this mathematical subtlety, we
point out a new moment in this concern.

Actually, the fact that we can define the atom (the system
$\mathcal{C}$) in the two ways, $e + p$ (the interacting
subsystems $\mathcal{E} + \mathcal{F}$) and $CM + R$ (the
noninteracting subsystems $\mathcal{A} + \mathcal{B}$),
distinguishes this issue as an issue falling within the context of
the question of what is system.

The method stemming from the decoherence theory may help in this
regard as pointed out in [5]. Apart from the mathematical
subtleties, one may note that the {\it separable} (Coulomb)
interaction [5] in the system $e + p$ allows re-definition of the
original pair of "particles" ($e$ and $p$) as the newly defined
(sub)systems ($CM + R$) of the one and the same composite
system--of the hydrogen atom. But, now, one may pose the question:
what actually {\it is} the system--the electron (or the proton) or
the relative particle (or the center-of-mass of the atom)? For an
isolated system $\mathcal{C}$, one would expect, at least in
principle, the full equivalence of the two possible divisions of
the composite system into subsystems. Needless to say, this way
posed, this question is the essence of the problem what is system.

As to the hydrogen atom, we know that the system $R$ appears
effectively as "system", not yet the electron. Actually, the
(internal atomic) energy we measure (by detecting the photons
emitted by the atom) and accessibility to measurement of the
relative position of the electron--as recently demonstrated  by
Maeda et al [4]--refer to the system $R$; the small numerical
margin for distinguishing $R$ from $e$ is just unimportant for our
discussion. Certainly, the accessible measurements (observables)
provide the (accessible) information about the "system" $R$, and
the fact that we have an information is at the hart of answering
the question of what is system.

Now, neglecting the fact that the atoms are open systems (cf.
Section 3), we may re-iterate our question: {\it if} the atom were
an isolated system, would we ever be able positively to answer the
question of what is system, or we would be able {\it equally} to
manipulate the electron (proton) states by LOCC? Bearing in mind
the (Coulomb) {\it interaction} (which bears separability [5, 7])
in $e + p$ system, one may expect entanglement of states of $e$
and $p$. Then, by neglecting the numerical indistinguishability of
the electron and $R$, one may wonder about the possibility to
manipulate the entanglement in $e + p$ by the proper operations
targeting the observables of $e$ and/or $p$; certainly, such
operations would require simultaneous operations on both $CM$ and
$R$.

From the mathematical point of view, the state of $CM + R$ system
is of the separable form $\vert \psi\rangle_{CM} \vert
\phi\rangle_R$, which in the position-representation becomes a
product of the "wave functions" of the form $\psi (\vec R_{CM})
\phi (\vec r_R)$; $\vec r_R \equiv \vec r_e - \vec r_p$, the
states $\phi (\vec r_R)$ representing the "stationary" states for
the atom. On the other side, due to the Coulomb interaction, the
states of the $e + p$ system are expected to bear entanglement,
which in the position representation obtains the form $\sum_i C_i
\Psi_i (\vec r_e) \chi_i (\vec r_p)$. The fact that we deal with
the unique composite system--the hydrogen atom--stems the
equality:

\bigskip

\begin{equation}
\psi (\vec R_{CM}) \phi (\vec r_R) = \sum_i C_i \Psi_i (\vec r_e)
\chi_i (\vec r_p)
\end{equation}

\bigskip

As to the dynamics of the atom, one may write the Hamiltonian of
the two divisions of the atom as follows:

\bigskip

\begin{equation}
\hat H = \hat T_{CM} \otimes \hat I_R + \hat I_{CM} \otimes (\hat
T_{R} + V(\hat r_R))
\end{equation}

\bigskip

\begin{equation}
\hat H = \hat T_{e} \otimes \hat I_p + \hat I_e \otimes \hat T_{p}
+ V_{Coul}(\hat{\vert\vec r_e} - \hat{\vec r_p\vert})
\end{equation}

\bigskip

\noindent for the pair $CM + R$ and $e + p$, respectively; by
$\hat T$ we denote the kinetic terms of the Hamiltonian, while
$V(\hat r_R) \equiv V_{Coul}(\hat{\vert\vec r_e} - \hat{\vec
r_p\vert})$. As it is apparent, the two divisions of the composite
system may be the basis of the different, {\it non-equivalent
quantum computations} generated by the one and the same
Hamiltonian $\hat H$. Actually, due to the lack of any
interaction (and consequently of entanglement) between $CM$ and
$R$, one can not expect any useful {\it quantum}
information/computation directly (i.e. without any external
action) to be performed by these subsystems. On the other side,
the expected entanglement between $e$ and $p$ may in principle be
useful for the information/computation processing. Therefore, the
hydrogen atom initially defined as the  $e + p$ system (of
mutually non-entangled, non-interacting subsystems $e$ and $p$)
still bears a "hidden" entanglement as well as the possibility of
performing the entanglement-based quantum computations.

Operationally, in the more general terms, our question reads:
along with accessibility of the observables (information) of e.g.
$\mathcal{A}$, we wonder about the possibility of the information
processing on the basis of accessibility of the observables of
e.g. the system $\mathcal{F}$. To this end, the approach of
Zanardi et al [9] that {\it stipulates} the "experimentally
accessible observables" seem to be an {\it ad hoc} not yet
necessarily the general answer to the question posed. The fact we
do not offer a definite answer to this question seems to be a
consequence of its deep connection with the fundamental problem of
what is system. Unfortunately, as we show in Section 3, there is
an even more limiting condition that has been neglected so far in
our discussion.

But the main observation may seem to be striking in the
information-theoretic context: even in a system as {\it simple} as
the (isolated) hydrogen atom, one can in principle perform the
information processing by manipulating entanglement in the
composite system yet dealing with the {\it non-interacting},
non-entangled subsystems. Needless to say, in order to do so,
experimenter should be able to operate in terms of the observables
of the "new" systems (cf. $\mathcal{E}$ and $\mathcal{F}$
above)--that still may be an open question of our considerations.
As a consequence, the LOCC referring to the pair $e + p$ seem just
to be the composite system operations relative to the pair $CM +
R$, and {\it vice versa}--which is a natural consequence of the
relativity [5] of the concept of physical system. Now, the
concepts of the "composite observables"--widely used and discussed
in the foundations of quantum mechanics and decoherence theory [2,
10, 11]--as well as the "local" observables become relative. E.g.,
the best known and studied "composite observables", the $CM$
position coordinates [1, 2, 11] relative to the system $e + p$,
become just the local observables relative to the $CM + R$ system.
Certainly, a definition of the observables does not yet imply the
measurement accessibility of the observables.

Finally, one may go even further in this concern by speculating in
the following way: what if instead of the bipartite system
$\mathcal{E} + \mathcal{F}$, one obtains a higher complexity
(coarse graining) of $\mathcal{C}$, such as the one formally
presented as follows: $\mathcal{A} + \mathcal{B} = \mathcal{C} =
\mathcal{M} + \mathcal{N} + \mathcal{P} \dots$? Or even more
interesting: what if an initially undivisable ("elementary")
"particle" can be decomposed following the recipes of the general
technique for defining "(sub)systems" [5]?

In conclusion to this section, we want to emphasize that
investigating the (as yet open) issue of "what is 'system'?" may
in principle help in finding the easier ways and methods for
manipulating quantum entanglement and for performing quantum
computation yet in the seemingly simple systems. It seems that
answering these questions (what is system, and which observables
can operationally, e.g. by LOCC, be accessed) should be answered
in parallel, and the outcome is not an easy matter to predict.
Yet, it seems we have already learnt: in general, there are not
{\it a priori} the "systems" (i.e. {\it a priory} inaccessible
observables of the (isolated) "system")--the lesson sounding very
much like the old lesson of Bohr (cf. Introduction) stemming from
quantum complementarity.

\bigskip

{\bf 3. The open systems limitations}

\bigskip

\noindent The hydrogen atom is an {\it open} system in interaction
with the quantum vacuum fluctuations. Without this interaction,
the stationary states of the atom would bear the full
stability--as expected solely from the Schrodinger equation for
the atom [12].

The presence of this environment (the vacuum fluctuations
$\mathcal{V}$) gives rise to both nonstationary character of the
"stationary" states of the atom and to the special status of its
ground state. More precisely: the system $R$ is in interaction
with $\mathcal{V}$, leaving yet the system $CM$ intact. The
environment-induced behavior of $R$ is then rather expected [1, 2,
3, 7] thus being the origin of the possibility to define $R$ as
"system". In other words: the environment $\mathcal{V}$ is
responsible for "accessibility" of $R$'s observables and the
related information about the system $R$, which  thus appears
effectively to be a "real" physical system accessible to
observation in a laboratory.

Certainly, this observation reinforces our main question: whether
or not we will ever be able operationally to manipulate the pair
$(e, p)$ and to employ their (expected) quantum entanglement?

Therefore, the discussion of Section 2 bears certain limitation:
it directly refers to the {\it isolated} systems, while its
relevance to the context of open systems will be  outlined in the
sequel.

\bigskip

{\bf 4. Extracting information about "subsystems"}

\bigskip

\noindent Due to the general rules of decoherence theory: the
environment selects the preferred states of the open system by
effectively forbidding their coherent (linear) superpositions [3].
For an independent observer, these states (that bear certain
robustness relative to the external actions) appear to be
"objective", thus in effect giving rise to a basis of defining
the open system.

{\it Prima facie}, it seems that the answer as to what is system
is already given and the relativity of "system" might seem to be
of the secondary importance. While the definition of "system" in
general is far from being complete [1, 5, 13], let us focus on
the task of extracting the information about the alternative
subsystems, again in terms of the hydrogen atom.

The measurements of e.g. the position coordinates of both $CM$ and
$R$ can directly lead to an information about the position of both
$e$ and $p$. Actually, by the use of the transformations {\it
inverse} to the well-known {\it canonical transformations} of
coordinates of $e$ and $p$:

\begin{equation}
\vec R_{CM} = (m_p \vec r_p + m_e \vec r_e)/ (m_p + m_e)
\end{equation}

\begin{equation}
\vec r_R = \vec r_e - \vec r_p
\end{equation}

\noindent one can calculate the values for $\vec r_e$ and $\vec
r_p$, in the full analogy with the classical system analysis.
Thus obtained information about the position of $e$ and $p$
refers equally to {\it both}, the isolated as well as the open
system.

However, as to the open systems, the operational use of such
information bears a subtlety to be emphasized. Actually, in order
to be able to exctract further information or to manipulate the
information obtained about $e$ and/or $p$, one should, in general,
be skilled enough to operate in the time intervals much shorter
than the decoherence time referring to the decoherence of $R$; the
decoherence induced by $\mathcal{V}$ (Section 3). Namely, the
decoherence of $R$ states inevitably affects the system $e + p$,
possibly giving rise to affecting the entanglement in $e + p$
system. And this is a general possible obstacle for the
operational information processing in the alternative subsystems
of an open composite system. On the other side, as to the isolated
systems--such as those dealt with in quantum information
theory--there seems to be {\it no such obstacles} for the
operational use of the relativity of "system" as distinguished in
Section 2.

There is another yet general notion and the possible obstacle to
our program of exctracting information about subsystems. In
general, the canonical transformations defining the different
divisions of a composite system into subsystems may include the
conjugate momenta of the subsystems, not only the position
variables as given in eq. (4) and (5)--cf. Appendix A. Then, as
pointed out in [5], such divisions of the composite system become
mutually {\it complementary}: due to incompatibility of the
position and momentum observables, the inverse transformations (in
analogy with eqs. (4) and (5)) can not be defined (due to the {\it
lack} of the {\it simultaneous} sharp values of the position and
the momentum observables). Therefore, in general, even for the
isolated composite system, one can not extract information about
the states and/or observables of the complementary subsystems of a
composite system.

However, if non-sharp (e.g. the mean) values of the position
and/or of the momentum observables might be useful, then it seems
that the case of the complementary subsystems reduces to the
non-complementary subsystems as discussed in Section 2. E.g. a
decay of an excited state of $R$ may uniquely determine the
initial (excited) state. Now, having this information, it is a
simple task to calculate the mean value of the position observable
$\hat {\vec r_R}$. By simultaneous measurement of $\hat {\vec
R_{CM}}$, one can directly obtain an approximate value of the
average position of both, $e$ and $p$, still with the standard
deviations not significantly exceeding the standard deviations of
$\hat {\vec r_R}$ and $\hat {\vec R_{CM}}$.

\bigskip

{\bf 5. Discussion and conclusion}

\bigskip

\noindent It is essential to re-emphasize: defining a
"(sub)system" assumes neither regropuing the subsystems nor a
redefinition of a unique subsystem's variables. Rather, it
assumes the canonical transformations that {\it couple the
variables of the different (sub)systems} [5, 13] thus, in
general, {\it not providing any intuitive contents} for the newly
defined (sub)systems even in the classical-physics context. To
this end, in order to circumvent the possible misunderstanding,
it is worth re-emphasizing that our considerations apply to the
variables transformations allowing a definition of the "new"
subsystems, not just giving rise to a definition of a new
composite observable of the composite system. To this end, as an
example we point e.g. to the observable defined as the sum
$\hat{\vec J} = \hat{\vec l} + \hat {\vec s}$, where $\hat{\vec
l}$ stands for the angular momentum and $\hat{\vec s}$ for the
spin of a particle; the observable $\hat{\vec J}$ is truly a
composite system observable {\it not yet defining the new
subsystems}.

As it is emphasized in [5], a division of a composite system gives
rise to {\it simultaneous} definitions of the subsystems not yet
allowing simultaneous observability of the subsystems belonging to
the different divisions. While this is a simple  consequence of
the canonical transformations (as defined in Appendix A) [5], this
fact bears a clear-cut information-theoretic aspect. Actually, the
observation of e.g. $\mathcal{A}$ represents a local operation
{\it relative} to the division $\mathcal{A} + \mathcal{B}$, while
the observation of $\mathcal{E}$ represents an operation exerted
on the composite system $\mathcal{A} + \mathcal{B}$--and therefore
can not be simultaneously performed. Certainly, this is another
subtlety of the problem at issue.

The coordinates-transformations-defined (sub)systems is virtually
a general method in physics, which makes our considerations to be
of interest for the {\it realistic} physical models. Here, we
shall outline just a few examples in this regard.

E.g., besides the hydrogen atom (Sections 2 and 4), we emphasize
the relevance of the contents of Section 2 for the widely used
method of redefinition of mutually interacting systems as a
collection of mutually non-interacting quasiparticles--e.g. the
"normal coordinates" decoupling the interacting harmonic
oscillators (cf. Appendix B) .

As another interesting issue, we emphasize the subject of the
molecules structure (a general issue of quantum chemistry), and
the problem of the macromolecules conformational transitions [14,
15, 16]. Actually, the general method of quantum chemistry reads:
a molecule can be modeled as a collection of the atomic nuclei
plus the collection of the atomic electrons $\mathcal{N} +
\mathcal{E}$. However, as it is generally treated in chemistry, a
molecule can be defined as a collection of mutually interacting
atoms (the system $\mathcal{A}$)--the interactions
(phenomenologically) described as the "chemical bonds" between the
atoms. In the context of our considerations, the composite system
$\mathcal{N} + \mathcal{E}$ is not quite equivalent with the
composite system $\mathcal{A}$, for the rather obvious reasons.
Actually, the adiabatic approximation valid for the system
$\mathcal{N} + \mathcal{E}$ is not valid  for the system
$\mathcal{A}$--the atomic mass ratio is a non-negligible fraction
of unity. Now, the adiabatic approximation--in its zeroth
order--gives rise to the separability of the subsystems
$\mathcal{N}$ and $\mathcal{E}$--the presence of quantum
entanglement in $\mathcal{N}$ + $\mathcal{E}$ system is considered
in the molecular spectroscopy theory as the domain of
non-applicability of the adiabatic approximation  [17]. On the
other side, the interactions between the atoms--to which the
adiabatic approximation does not apply--should provide
entanglement in the system of the atoms. Similarly, the externally
induced conformational transitions [14, 15, 16] of the molecules
[18, 19] sound different in terms of the two divisions
[16]--$\mathcal{N}$ + $\mathcal{E}$ and $\mathcal{A}$--of a
molecule, that is also characteristic for our considerations.
Being the many-particle systems, the molecules exhibit virtually
the general applicability of our considerations to the {\it
realistic} physical models.

A slight limitation of our considerations may be the fact that
our analysis primarily refers to the continuous observables.
However, the progress in the continuous-variables-based quantum
information processing (cf. e.g. [20, 21]) encourages the
applicability of our results to the general quantum
information/computation issues.

Interestingly enough, the method for defining new subsystems may
also be considered as a method for combating decoherence in the
quantum hardware. The separable states of a system (e.g. of $CM +
R$, cf. Section 2) may (and sometimes do) appear as a consequence
of decoherence. If so, the introduction of the new yet entangled
subsystems may {\it directly} help in principle in avoiding
entanglement in the composite system.

In conclusion, we want to stress: {\it even a seemingly simple yet
composite quantum system of mutually nonentangled, noninteracting
particles may bear (a "hidden") entanglement as well as the
possibility of performing the nonequivalent computations, relative
to the different possible divisions of the composite system into
subsystems}. This lesson justifies the following conclusion: a
{\it quantum} system is not {\it a priory} given as well as
accessibility of its observables. The method(s) (yet fully to be
formulated) in this regard may in principle be applied to both
isolated as well as to the open quantum systems--that also may be
a general method for avoiding decoherence in the quantum hardware.

\bigskip

{\bf Appendix A}

\bigskip

\noindent By "canonical transformations" we assume the standard
coordinate transformations within the Hamiltonian formalism of
classical mechanics. Their quantum mechanical counterparts
directly follow due to the procedure of quantization of the
classical variables.

E.g., let us assume that a composite system is defined by the two
subsystems, $\mathcal{A}$ and $\mathcal{B}$, each subsystem being
defined by the proper degrees of freedom and the corresponding
conjugate momenta--$\{x_{Ai}, p_{Ai}\}$ and $\{X_{Bj}, P_{Bj}\}$,
respectively. Then, the transformations are defined formally as
follows:

\begin{equation}
\xi_{Ek} = \xi_{Ek} (x_{Ai}, p_{Ai}; X_{Bj}, P_{Bj}), \quad
\pi_{Ek} = \pi_{Ek} (x_{Ai}, p_{Ai}; X_{Bj}, P_{Bj})
\end{equation}

\begin{equation}
\Xi_{Fk} = \Xi_{Fk} (x_{Ai}, p_{Ai}; X_{Bj}, P_{Bj}), \quad
\Pi_{Fk} = \Pi_{Fk} (x_{Ai}, p_{Ai}; X_{Bj}, P_{Bj})
\end{equation}

\noindent thus introducing the new "subsystems", $\mathcal{E}$ and
$\mathcal{F}$, respectively to eqs. (6), (7), still allowing
redefinition of the system Hamiltonian function:

\begin{equation}
H = H_A  +  H_B  +  H_{AB}
\end{equation}

\begin{equation}
H = H_E  +  H_F  +  H_{EF}.
\end{equation}

\noindent where e.g. $H_{AB}$ describes the interaction between
the subsystems $\mathcal{A}$ and $\mathcal{B}$.

It is worth emphasizing: the canonical transformations (6), (7)
{\it substantially redefine} the composite system due to, in
general, dependence of the degrees of freedom e.g. of
$\mathcal{E}$ of both, degrees of freedom {\it and} the conjugate
momenta of {\it both} systems, $\mathcal{A}$ and $\mathcal{B}$;
being linear, these transformations allow the transformations {\it
inverse} to (6), (7). In the other words: the new "subsystems",
$\mathcal{E}$ and $\mathcal{F}$, need {\it not} bear any {\it
intuitive} contents as the physical systems.

\bigskip

{\bf Appendix B}

\bigskip

\noindent A linear interaction of the two harmonic oscillators of
the general form $\hat H_{12} = C \hat x_1 \otimes \hat x_2$
allows a redefinition of the composite system ${1 + 2}$ in terms
of mutually noninteracting harmonic oscillators, $Q_1 + Q_2$.
Actually, the "normal coordinates" $\hat q_i, i = 1, 2$, as
defined by the following expression (a special case of eqs. (6),
(7)):

\begin{equation}
\hat x_1 = 2^{-1/2} (\hat q_1 + \hat q_2), \quad \hat x_2 =
2^{-1/2} (\hat q_1 - \hat q_2),
\end{equation}

\noindent that define the new subsystems, $Q_i, i = 1, 2$, give
rise to the lack of any interaction of the new subsystems: $\hat
H_{Q_1Q_2} = 0$ [17].

 As distinct from the pair of microscopic
oscillators--that is typical for the quantum information
theory)--let us assume that one oscillator (e.g. the oscillator 2)
in the pair is a "macroscopic" system. Then the linear coupling of
the position observable $\hat x_1$ of the "microscopic" and the
center of mass observable $\hat X_{CM2}$ of the macroscopic
oscillator may give rise to decoherence of the position states of
the microscopic oscillator, $\hat x_1$ [22]. Certainly, the
oscillators ($Q_1, Q_2$) described by the "normal coordinates"
remain decoupled [17] thus {\it not} providing any entanglement or
decoherence in the pair ($Q_1, Q_2$).
\bigskip

\centerline{\bf Acknowledgements}

\bigskip

The work on this paper was financially supported by the Ministry
of Science and Environmental Protection, Serbia, under contract no
141016.

\bigskip

\centerline{\bf Bibliography}

\bigskip

[1] W.H. Zurek, Prog. Theor. Phys. {\bf 89}, 281 (1993)

[2] D. Giulini, E. Joos, C. Kiefer, J. Kupsch, I.-O. Stamatescu
and H.D. Zeh, "{\it Decoherendce and the Appearance of a Classical
World in Quantum Theory}", Springer, Berlin, 1996

[3] W.H. Zurek, Phys. Today {\bf 48}, 36 (1991)

[4] H. Maeda, D.V.L. Norum, T.F. Gallagher, Science {\bf 307},
1757 (2005)

[5] M. Dugi\' c, J. Jekni\' c, {\it Int. J. Theor. Phys.} (in
press; online first: DOI 10.1007/s10773-006-9186-0)

[6] W.H. Zurek, Phys. Rev. D {\bf 26}, 1862  (1982)

[7] M. Dugi\' c, Phys. Scr. {\bf 56}, 560 (1997)

[8] M. Dugi\' c, M.M. \' Cirkovi\' c, Phys. Lett. A {\bf 302}, 291
(2002)

[9] P. Zanardi, D. Lidar, and S. Lloyd, Phys. Rev. Lett. {\bf 92},
060402 (2004)

[10] B. d'Espagnat, "{\it Conceptual Foundations of Quantum
Mechanics}", Benjamin, Reading, MA, 1971

[11] R. Omnes, "{\it The Interpretation of Quantum Mechanics}",
Princeton University Press, Princeton, 1994

[12] R. Graham, M. Miyazaki, Phys. Rev. A {\bf 53}, 2683 (1996)

[13] M. Dugi\' c, eprint arXiv quant-ph/9903037

[14] C. Levinthal, J. Chem. Phys. {\bf 65}, 44 (1968).

[15] K.A. Dill, S. Chan, Nature Struct. Biol. {\bf 4}, 10 (1997)

[16] D. Rakovi\' c, M. Dugi\' c, M. Plav\v si\' c, Materials
Science Forum,  {\bf 453-454}, 521 (2004)

[17] L.A. Gribov, "{\it From Spectra Towards the Theory of the
Chemical Reactions}" (in Russian), Russian Academy of Sciences,
Moscow, 2001

[18] T. Kanai, S. Minemoto, H. Sakai, Nature {\bf 435}, 470 (2005)

[19] A. Lendlein, H. Jiang, O. Junger, R. Langer, Nature {\bf
434}, 879 (2005)

[20] X. Jia et al, Phys. Rev. Lett. {\bf 93}, 250503 (2004)

[21] L. Lamata, J. Leon, J. Opt. B: Quantum Semiclass. Opt. {\bf
7}, 224 (2005)

[22] W.H. Zurek et al, Phys. Rev. Lett. {\bf 70}, 1187 (1993)

\end{document}